\documentclass[11pt]{article}
\usepackage{graphicx}
\begin{document}
\title{\textbf{\textsf{Wormholes supported by phantom-like modified Chaplygin gas} }}
\author{ Mubasher Jamil\footnote{mjamil@camp.edu.pk},\ \ M. Umar Farooq
and Muneer Ahmad Rashid\footnote{muneerrshd@yahoo.com}
\\ \\
\small Center for Advanced Mathematics and Physics, National University of Sciences and Technology\\
\small E\&ME Campus, Peshawar Road, Rawalpindi, 46000, Pakistan \\
} \maketitle
\begin{abstract}
We have examined the possible construction of a stationary,
spherically symmetric and spatially inhomogeneous wormhole spacetime
supported by the phantom energy. The later is supposed to be
represented by the modified Chaplygin gas equation of state. The
solutions so obtained satisfy the flare out and the asymptotic
flatness conditions. It is also shown that the averaged null energy
condition has to be violated for the existence of the wormhole.
\end{abstract}
\large
\section{Introduction}
One of the most exotic geometries that arise as solutions of
Einstein field equations is the wormhole. A typical two mouth
wormhole connects two arbitrary points of the same spacetime or two
distinct spacetimes. One observes that any typical Schwarzschild
spacetime contains a singularity at $r=0$ making it geodesically
incomplete. Ellis \cite{ellis} first observed that the coupling of
the geometry of spacetime to a scalar field can produce a static,
spherically symmetric, geodesically complete and horizonless
spacetime and thus termed it as a `drainhole' that could serve as
tunnel to traverse particles from one side to the other. Later on
Morris and Thorne \cite{morris,morris1} proposed that wormholes
could be thought of (imaginary) time machines that could render
rapid interstellar travel for human beings. While a black hole
possesses single horizon which forbids two way travel (in and out)
of the black hole but this problem does not arise in the absence of
horizon for a wormhole. Unfortunately the existence of wormholes
require the violation of the most cherished energy conditions of
general relativity (null, weak, strong and dominant) which are in
fact satisfied by any normal matter or energy \cite{picon}. In
particular, matter violating null energy condition is called `exotic
matter' \cite{christian}. Later it was  proposed that wormholes
could be constructed with arbitrary small quantities of exotic
matter \cite{kuhfittig1,visser}. A commonly known form of matter
violating these energy conditions is dubbed as `phantom energy'
characterized by the equation of state (EoS) $p=\omega\rho$, where
$p$ and $\rho$ are respectively, the pressure and the energy density
of the phantom energy, with $\omega<-1$. The existence of this
matter remains hypothetical but the astrophysical observations of
supernovae of type Ia and cosmic microwave background have suggested
the presence of phantom energy in our observable universe
\cite{caldwell,babichev}. It can exhibit itself as a source that can
induce an acceleration in the expansion of the universe. The typical
size of a wormhole can be of the order of the Planck length but it
can be stretched to a larger size if it is supported by exotic
phantom like matter \cite{lobo1,diaz1}. The accretion of phantom
energy can increase the mass and size of the wormhole and hence
guarantee the stability of the wormhole \cite{diaz,jamil}. The
astrophysical implications of wormholes are not exactly clear but it
is suggested that some active galactic nuclei and other galactic
objects may be current or former entrances to wormholes
\cite{kardashev}. It has been predicted that wormholes can also
produce gravitational lensing events \cite{dey}. Since wormholes are
horizonless, they can avoid undergoing any process of decay like
Hawking evaporation and hence can survive over cosmological times.
But a wormhole may form a black hole with a certain radial magnetic
field (a form of magnetic monopole) if it accretes normal matter and
consequently loses its structure.

Earlier, Rahaman et al \cite{rahaman} investigated the evolution of
wormhole using an averaged null energy condition (ANEC) violating
phantom energy and a variable EoS parameter $\omega(r)$. We here
investigate the same problem using a more general EoS for the
pressure density namely the modified Chaplygin gas. It is well-known
that the wormhole spacetime is inhomogeneous and hence requires
inhomogeneous distribution of matter. This can be made by
introducing two different pressures namely the radial and the
transverse pressure. Our analysis shows that the parameters adopted
in the equation of state for phantom energy have to be tuned such
that the radial pressure becomes negative in all directions and for
all radial distance. This result turns out to be consistent with
Sushkov \cite{sushkov}.

The paper is organized as follows: In the second section, we have
modeled the field equations for the wormhole spacetime and proposed
the methodology that is adopted in the later sections. Next, we have
investigated the behavior of energy condition ANEC for all the
wormhole solutions obtained. Finally, the last section is devoted
for the conclusion and discussion of our results.

\section{Modeling of system}

We start by assuming the static, stationary, spherically symmetric
wormhole spacetime specified by (in geometrized units $G=1=c$):
\begin{equation}
ds^2=-e^{2f(r)}dt^2+\left(1-\frac{b(r)}{r}\right)^{-1}dr^2+r^2(d\theta^2+\sin^2\theta
d\phi^2).
\end{equation}
Here $f(r)$ is the `gravitational potential function' while $b(r)$
is called `shape function' of the wormhole (see Ref.
\cite{kuhfittig22} for the consistent derivation of the above
metric). The radial coordinate $r$ ranges over $[r_o,\infty)$ where
the minimum value $r_o$ corresponds to the radius of the throat of
the wormhole. If $b(r)=2m(r)$, the later being the mass, then Eq.
(1) represents a `dark energy star' which may arise from a density
fluctuation in the Chaplygin gas cosmological background
\cite{lobo,andrew}. Note that $b(r=r_o)=r_o$ corresponds to the
spatial position of the wormhole throat. We shall, in this paper,
assume $f(r)=$ constant for the convenience of our calculations.
This choice, as a special case, is also physically motivated and
makes the time traveler to feel zero tidal force near the wormhole
\cite{rahaman,thorne}. A wormhole with small $|f^\prime(r)|$ in the
vicinity of the throat is likely to be traversable in the sense of
having low tidal forces. It also makes the wormhole to be
horizon-free.

We take the inhomogeneous phantom energy which is specified by the
stress energy tensor: $T_{00}=\rho$, $T_{11}=p_r$,
$T_{22}=T_{33}=p_{t}$. Here $p_r$ and $p_{t}$ are, respectively, the
radial and transverse component of the pressure while $\rho$ is the
energy density of the phantom energy. It represents a perfect fluid
(which is homogeneous and isotropic) if $T_{11}=T_{22}=T_{33}=p_r=p$
\cite{kuhfittig}. Note that in the stellar evolution, the difference
$p_r-p_{t}$ creates a surface tension inside star which makes it
anisotropic. This feature is generically found in more compact stars
like neutron and quark stars \cite{jamil1}, contrary to normal stars
which are majorally supported by radial pressure only against
gravity.

The Einstein field equations $(G_{\alpha\beta}=8\pi
T_{\alpha\beta})$ for the metric (1) are
\begin{eqnarray}
\frac{b^{\prime }(r)}{ r^{2}}&=&8\pi\rho (r),\\
 -\frac{b}{r^{3}}&=&8\pi
p_r(r),\\ \left(1-\frac{b}{r}\right)\left[\frac{-b^{\prime
}r+b}{2r^{2}(r-b)}\right]&=&8\pi p_{t}(r).
\end{eqnarray}
The energy conservation equation is obtained from
$T^{\alpha\beta}_{;\alpha}=0$, which gives
\begin{equation}
p_r^{\prime }+\frac{2}{r}p_r-\frac{2}{r}p_{t}=0.
\end{equation}
This equation can be considered as the hydrodynamic equilibrium
equation for the exotic phantom energy supporting the wormhole.

 Eq. (2) can be written in the form
\begin{equation}
\frac{db}{dr}=8\pi r^{2}\rho.
\end{equation}
Let us choose the modified Chaplygin gas (MCG) EoS for the radial
pressure \cite{jamil2}
\begin{equation}
p_r(r)=A\rho(r) -\frac{B}{\rho(r) ^{\alpha }}.
\end{equation}
Here $A$, $B$ and $\alpha$ are constant parameters. The MCG best
fits with the $3-$year WMAP and the SDSS data with the choice of
parameters $A=-0.085$ and $\alpha=1.724$ \cite{lu} which are
improved constraints than the previous ones $-0.35<A<0.025$
\cite{jun}. Recently it is shown that the dynamical attractor for
the MCG exists at $\omega=-1$, hence MCG crosses this value from
either side $\omega>-1$ or $\omega<-1$, independent to the choice of
model parameters \cite{jing}. Generally, $\alpha$ is constrained in
the range $[0,1]$ but here we are assuming it to be a free parameter
which can take values outside this narrow range, for instance
$\alpha=-1$ as considered below. This later choice $\alpha<0$ makes
Eq. (7) a combination of a barotropic and a polytropic equation of
state.

Let us take the transverse pressure $p_{t}$ to be linearly
proportional to the radial pressure $p_r$ as
\begin{equation}
p_{t}=np_r,
\end{equation}
where $n$ is a non-zero constant. Thus $p_r$ is restricted to
satisfy Eq. (7) for a given $\rho$ while $p_t$ is arbitrary in
nature due to free parameter $n$. Using Eq. (8) in (5), we obtain
\begin{equation}
p_r^{\prime }+\frac{2}{r}p_r-\frac{2n}{r}p_r=0,
\end{equation}
which gives
\begin{equation}
p_r=Cr^{2(n-1)}.
\end{equation}
Here $C$ is a constant of integration. Since the wormhole is
supported by a negative pressure inducing exotic phantom energy, it
yields $p_r<0$ if $C<0$ and $1<n<\infty$ in order to obtain finite
negative radial pressure. Consequently $p_{t}<0$ if $0<n<\infty$.
Using Eq. (10) in (7), we have
\begin{equation}
A\rho -\frac{B}{\rho ^{\alpha }}=Cr^{2(n-1)},
\end{equation}
which can be written as
\begin{equation}
A\rho ^{\alpha+1 }-Cr^{2(n-1)}\rho ^{\alpha} -B=0.
\end{equation}
Note that Eq. (12) is a polynomial equation of degree $\alpha+1$ in
variable $\rho$, which does not yield solutions for any arbitrary
$\alpha$. We shall, henceforth, solve Eq. (12) for specific choices
like $\alpha=-1$, $0$ and $1$. We shall further employ the following
conditions on our solutions given below \cite{rahaman1}:

1. The potential function $f(r)$ must be finite for all values of
$r$ for the non-existence of horizon. In our model, this condition
is trivially satisfied since $f(r)$ is taken to be a finite constant
throughout this paper.

2. The shape function $b(r)$ must satisfy $b'(r=r_o)<1$ at the
wormhole throat with radius $r_o$, the so-called flare-out
condition.

3. Further $b(r)<r$ outside the wormhole's throat $r>r_o$. This
condition is a direct consequence of the flare-out condition.

4. The spacetime must be asymptotically flat i.e.
$b(r)/r\rightarrow0$ for $|r|\rightarrow\infty$.

Now we shall consider the three cases for different choices of
parameter $\alpha$:

\textbf{Case-a:} If $\alpha =0,$ then (12) gives
\begin{equation}
\rho =\frac{B}{A}+\frac{C}{A}r^{2(n-1)}.
\end{equation}
Using Eq. (13) in (6), we get
\begin{equation}
b(r)=\frac{8\pi}{A} \left[\frac{Br^3}{3}+
\frac{Cr^{2n+1}}{2n+1}\right]+C_1.
\end{equation}
Here $C_1$ is a constant of integration. Now
$\frac{b(r)}{r}\rightarrow0$ as $|r|\rightarrow\infty$ if $n=1$ and
$B=-C$. But here $b(r)=$ constant and hence gives $\rho=0$ which is
an acceptable solution and represents vacuum (empty space-time)
outside the wormhole throat. This corresponds to vanishing pressures
i.e. $p_r=p_{t}=0$. This vacuum solution requires $C=0$ which in
turn leads to $B=0$. Note that condition (4) can also be met if only
$B=0$ and $n<0$. In figures 1 to 4, we have plotted the ratio
$b(r)/r$ against the parameter $r$. The Fig. 1 shows that the ratio
declines as $r\rightarrow\infty$, although $r$ is restricted to a
certain range. The parameter $A$ can assume the value in the range
$-0.35\leq A\leq 0.025$ \cite{wu2}, we choose $A=0.025$ for our
work.

Further, flare-out condition (2) implies
\begin{equation}
b'(r_o)=\frac{8\pi C}{A} r_o^{2n}<1,
\end{equation}
which gives an upper limit on the size of throat's radius as
\begin{equation}
r_o<\left(\frac{A}{8\pi C}\right)^\frac{1}{2n}.
\end{equation}
This requires both $A>0$ and $C>0$. The throat's radius can be
obtained by solving $b(r_o)=r_o$ which gives
\begin{equation}
r_o=\left[\frac{A(2n+1)}{8\pi C}\right]^\frac{1}{2n}.
\end{equation}
This quantity is positive if $A/C>0$ and $2n+1>0$ or $A/C<0$ and
$2n+1<0$. Similarly, condition (3) translates into
\begin{equation}
r<\left[\frac{A(2n+1)}{8\pi C}\right]^\frac{1}{2n}, \ \ r>r_o.
\end{equation}
Note that conditions (2) and (3) are satisfied if $C_1=0$.

\textbf{Case-b:} If $\alpha =1,$ then (12) gives
\begin{equation}
A\rho ^{2}-Cr^{2(n-1)}\rho -B=0,
\end{equation}
which is quadratic in $\rho $ and gives two roots of the form
\begin{equation}
\rho_{\pm} =\frac{Cr^{2(n-1)}\pm \sqrt{C^{2}r^{4(n-1)}+4AB}}{2A}.
\end{equation}
These roots are real-valued if the quantity inside the square root
is positive while the roots will be repeated if it is zero and
complex valued otherwise. We next determine the shape function
$b(r)$ corresponding to these roots by substituting Eq. (20) in (6)
to get
\begin{eqnarray}
b_\pm(r)&=&\pm\frac{4\pi
r(\pm Cr^{2n}+r^2\sqrt{4AB+C^2r^{4(n-1)}})}{A(1+2n)}+C_{2\pm}\nonumber \\
&\;&\pm[2B\pi
r^7\sqrt{1+\frac{4ABr^{4(1-n)}}{C^2}}\sqrt{4AB+C^2r^{4(n-1)}}\Gamma\left(\frac{1+2n}{4(1-n)}\right)\nonumber\\
&\;&\times
_{2}F_{1}^\prime\left(\frac{5-2n}{4(1-n)},\frac{1}{2},\frac{9-6n}{4(1-n)},-\frac{4ABr^{4(1-n)}}{C^2}\right)]/[(n-1)(4ABr^4+C^2r^{4n})].
\end{eqnarray}
Here $C_{2\pm}$ are two constants of integration whereas
$_{2}F_{1}^\prime$ is the regularized hyper-geometric function.
Figures 2 and 3 show the ratios $b_+(r)/r$ and $b_-(r)/r$ versus
$r$, respectively. Both ratios decline for large values of $r$ and
approach zero, satisfying the asymptotic flatness condition for
specific choice of the parameters.

\textbf{Case-c:} If $\alpha =-1,$ then (12) yields
\begin{equation}
\rho =\frac{C}{A-B}r^{2(n-1)},
\end{equation}
Use of this in (6) enables us to write
\begin{equation}
b(r)=\frac{8\pi C}{2(n+1)(A-B)}r^{2(n+1)}.
\end{equation}
In figure 4, the ratio $b(r)/r$ is plotted against $r$, showing its
convergence to zero. Further, condition (2) implies
\begin{equation}
b'(r_o)=\frac{8\pi C}{A-B}r_o^{2n+1}<1,
\end{equation}
which gives the maximum size of the wormhole's throat
\begin{equation}
r_o<\left(\frac{A-B}{8\pi C} \right)^{\frac{1}{2n+1}}.
\end{equation}
In other words, the throat's radius is given by
\begin{equation}
r_o=\left[\frac{2(n+1)(A-B)}{8\pi C}\right]^{\frac{1}{2n+1}}.
\end{equation}
It requires either $A-B>0$, $C>0$ and $n>-1$ or $A-B<0$ and $C<0$.
This later choice of parameters is consistent with the ones that are
required for $p_r<0$ in Eq. (10).

As we discussed earlier, the relativistic energy conditions are
satisfied by ordinary classical matter but there are some physical
processes where these conditions are violated. For example, for a
black hole evaporation caused by the emission of Hawking radiation
\cite{hawking}. The quantized fields in the surrounding of black
hole produce massive particles carrying positive energy density. Due
to energy conservation in the whole process, the negative energy
density is added to the total energy density of the black hole.
Consequently the black hole loses mass and its horizon shrinks. The
energy conditions are also violated when an electromagnetic wave is
squeezed resulting in the energy density of the wave to become
negative, zero and positive at certain wavelengths but the averaged
energy density of the wave remains positive \cite{thorne}. One
observes that the notion of violation of energy conditions is quite
ubiquitous at the quantum scale. As the quantum effects allow for a
localized violation of energy condition, there is a limit to an
extent by which these conditions can be violated globally. In this
connection, the `averaged null energy condition' (ANEC) is specified
which states that \cite{poisson}
\begin{equation}
\int\limits_\gamma T_{\alpha\beta}k^\alpha k^\beta d\lambda\geq0.
\end{equation}
Here $T_{\alpha\beta}$ is the stress energy tensor, $k^\alpha$ is
the future directed null vector, $\gamma$ is the null geodesic and
$\lambda$ is the arc-length parameter. In other words, the integrand
must be positive. In an orthonormal frame of reference, we have
$k^{\hat{\alpha}}=(1,1,0,0)$, so that
$T_{\hat{\alpha}\hat{\beta}}k^{\hat{\alpha}}k^{\hat{\beta}}=\rho+p_r$.
We here adopt the ANEC integral from Visser et al \cite{visser} to
analyze its violation in our model:
\begin{equation}
I=\oint(\rho+p_r)dV=2\int_{r_o}^\infty(\rho+p_r)4\pi r^2dr.
\end{equation}
The above integral is called the `volume integral quantifier'
\cite{lobo3}. It is obvious that the above integral becomes negative
if $\rho+p_r<0$, the violation of null energy condition (see also
\cite{subenoy}). The violation of ANEC is the requirement for any
phantom matter and stability of a wormhole. Now we shall take
different $\rho$ and $p_r$ calculated in each of the above cases to
evaluate $I$. Our aim will be to find conditions under which $I<0$.
We shall also consider the case of $I\rightarrow0$ which suggests
the construction of wormhole with arbitrarily small amounts of a
phantom energy.

\textbf{Case-a} Using Eqs. (7) and (13) in (28), we obtain
\begin{equation}
I=\left.\frac{8\pi C(1+A)}{A(1+2n)}r^{1+2n}\right\vert_{r_o}^\infty.
\end{equation}
Note that the above integral gives a finite value if $n<-1/2$. Hence
we obtain
\begin{equation}
I=-(1+A)\left[\frac{A(1+2n)}{8\pi C}\right]^{\frac{1}{2n}}.
\end{equation}
Moreover, the above integral $I<0$ if $1+A>0$ or $A>-1$. Further,
$A/C<0$ which implies either (1) $C<0$ and $A>0$ or (2) $C>0$ and
$A<0$. Again the former case (1) is consistent with $p_r<0$. Also
$I\rightarrow0$, if either $A\rightarrow0$ or $n\rightarrow-1/2$.

\textbf{Case-b} Using Eqs. (7) and (21) in Eq. (28), we get
\begin{eqnarray}
I_\pm&=&\frac{4\pi
r((1+2A)Cr^{2n}\pm r^2\sqrt{4AB+C^2r^{4(n-1)}})}{A(1+2n)}\nonumber \\
&\;&\mp[32(n-1)B\pi
r^7\sqrt{1+\frac{4ABr^{4(1-n)}}{C^2}}\sqrt{4AB+C^2r^{4(n-1)}}\\
&\;&\times
\left._{2}F_{1}^\prime\left(\frac{5-2n}{4(1-n)},\frac{1}{2},\frac{9-6n}{4(1-n)},-\frac{4ABr^{4(1-n)}}{C^2}\right)]/[(2n+1)(2n-5)(4ABr^4+C^2r^{4n})]\right\vert_{r_o}^\infty.\nonumber
\end{eqnarray}
The ANEC is violated for particular choice of parameters like
$B=-3$, $C=-2$ and $n=3$. Note that we have assumed $C_{2\pm}=0$ for
the convenience of our calculations. Under this choice of
parameters, the two integrals $I_{\pm}$ will be finite. Also figures
(5) and (6) show the behavior of the integrals $I_+<0$ and $I_-<0$,
respectively. The plots suggest that the two integrals $I_{\pm}$
tend to infinity for large $r$, so that an infinite amount of ANEC
violating matter is necessary to sustain these geometries, which is
a problematic issue. However this problem can be evaded by
considering a matching to an exterior vacuum solution which gives a
thin-shell wormhole solution \cite{lobo2,eiroa,subenoy}. Further,
the case of $I_\pm\rightarrow0$ arises if $n\rightarrow1$,
$A\rightarrow-1/2$ and $C^2+4AB\rightarrow0$.

\textbf{Case-c} Making use of Eqs. (7) and (22) in (28) yields
\begin{equation}
I=\left.\frac{8\pi C(1+A-B)
r^{1+2n}}{(A-B)(1+2n)}\right\vert_{r_o}^\infty.
\end{equation}
The above integral is finite if $n<-1/2$. Therefore we obtain
\begin{equation}
I=\frac{1+A-B}{1+2n}\left(\frac{A-B}{8\pi C}\right)^{\frac{1}{2n}}.
\end{equation}
Further, the ANEC is violated $I<0$ if either $C<0$ or $A-B<0$. The
wormhole is supported by arbitrary small amount of phantom energy if
$A-B\rightarrow0$.

\section{Conclusion and discussion}

In this paper, we have derived three solutions of wormhole by
obtaining different forms of $b(r)$. This is carried out by
employing the modified Chaplygin gas for the pressure and using
three specific values of the parameter $\alpha$. It needs to be
mentioned that other values of $\alpha$ either don't yield any
$b(r)$ or if it does exist than the stability conditions 1 to 4 are
not verified. Hence we have restricted ourselves to these specific
cases as shown in the figures as well. The solutions so obtained
also satisfy the stability conditions. The pressure and the
corresponding energy density obtained in each case, violate the null
energy condition $\rho+p_r<0$ and hence the averaged null energy
condition is also violated. These conditions need to be violated for
the existence of any wormhole solution.

We performed the analysis by taking the wormhole spacetime to be
inhomogeneous and anisotropic with non-vanishing transverse
pressure. The spacetime needs to be anisotropic as it was found that
considering an isotropic pressure $p_r=p_t=p$, for $f(r)$ to be
finite, one cannot construct asymptotically flat traversable
wormhole \cite{lobo1}. In our work, we represented the radial
pressure by the modified Chaplygin gas and the transverse pressure
to be linearly proportional to the radial one. The MCG has phantom
nature with negative pressure. Earlier, Lobo \cite{lobo3} studied
the Chaplygin traversable wormhole and concluded that the Chaplygin
gas needs to be confined around the wormhole throat neighborhood.
That work was later extended in \cite{subenoy} using the modified
Chaplygin gas and it was deduced that modified Chaplygin wormholes
may occur naturally and could be traversable. Our results are in
conformity with their results since our solutions meet the criteria
of wormhole stability and traversability.

In a recent paper, Gorini et al \cite{gorini} have presented an
interesting theorem which states that in a static spherically
symmetric spacetime filled with the phantom Chaplygin gas, the
scalar curvature becomes singular at some finite value of the radial
coordinate $r$ and henceforth the spacetime is not asymptotically
flat. This result apparently forbids the existence of wormholes
which are required to be non-singular. The theorem is based on the
assumptions of homogeneity and isotropy of the spacetime. In case of
anisotropy ($p_t\neq0$), the above theorem is not applicable and
wormhole spacetime appears naturally. For an isotropic and
homogeneous spacetime filled with phantom Chaplygin gas, the
asymptotic flatness can be achieved by cutting the spacetime at some
spatial position $r=R$ and glued with a vacuum spacetime, in
particular, Schwarzschild exterior spacetime can be utilized
\cite{lobo3}.

\subsubsection*{Acknowledgment}
One of us (MJ) would like to thank A. Qadir and M. Akbar for useful
discussions during this work. We would also acknowledge anonymous
referees for their useful criticism on this work.

\pagebreak
\newpage
\begin{figure}
\includegraphics{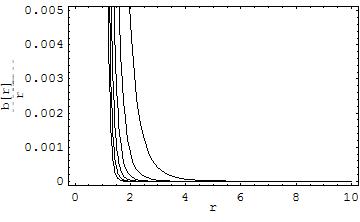}\\
\caption{The ratio $b(r)/r$ is plotted against $r$ with $C=3$ and
for different values of $n=-2,-2.5,-3,-3.5,-4,-4.5$ which correspond
to curves in right to left order.}
\end{figure}
\begin{figure}
\includegraphics{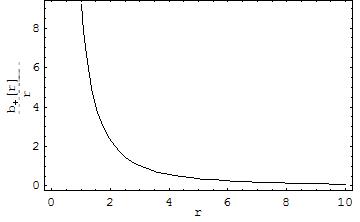}\\
\caption{The ratio $b_+(r)/r$ is plotted against $r$. The parameters
are fixed at $B=6, C=7$ and $n=3$.}
\end{figure}
\begin{figure}
\includegraphics{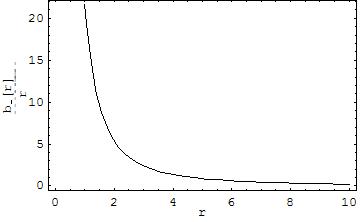}\\
\caption{The ratio $b_-(r)/r$ is plotted against $r$. The parameters
are fixed at $B=-2, C=7$ and $n=3$.}
\end{figure}
\begin{figure}
\includegraphics{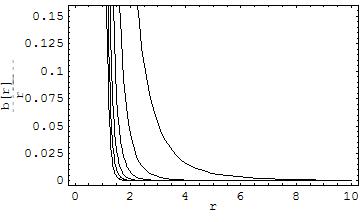}\\
\caption{The ratio $b(r)/r$ is plotted against $r$ for different
values of $n=-3,-4,-5,-6,-7,-8$. The parameters are taken $B=1$,
$A=-5$ and $C=2$ which correspond to curves in right to left order}.
\end{figure}
\begin{figure}
\includegraphics{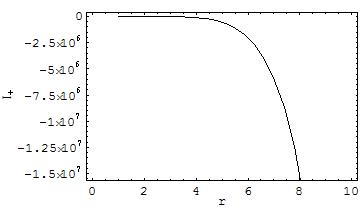}\\
\caption{The ANEC integral $I_+$ is plotted against $r$. The
apparent negative values of $I_+$ show the violation of the ANEC
condition. The parameters are chosen as $B=-3$, $C=-2$ and $n=3$}
\end{figure}
\begin{figure}
\includegraphics{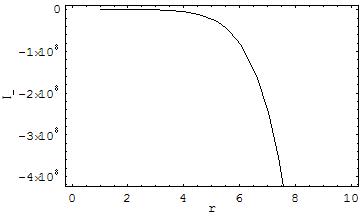}\\
\caption{The ANEC integral $I_-$ is plotted against $r$. The
apparent negative values of $I_-$ show the violation of the ANEC
condition. The choice of parameters is the same as in figure 5.}
\end{figure}

\end{document}